\documentclass[]{fairmeta}
\usepackage{amssymb,amsthm,amsmath}
\usepackage{mathtools}
 \usepackage{xspace}

\renewcommand{\phi}{\varphi}

\usepackage{algorithm}
\usepackage{algorithmic}
\usepackage{float}
\usepackage{enumitem}
\usepackage{newtxtext}
\usepackage{newtxmath}

\title{Idea2Story: An Automated Pipeline for Transforming Research Concepts into Complete Scientific Narratives}

\author{Tengyue Xu*}
\author{Zhuoyang Qian*}
\author{Gaoge Liu*}
\author{Li Ling*}
\author{Zhentao Zhang*}
\author{Biao Wu*}
\author{Shuo Zhang}
\author{Ke Lu}
\author{Wei Shi}
\author{Ziqi Wang}
\author{Zheng Feng}
\author{Yan Luo}
\author{Shu Xu}
\author{Yongjin Chen}
\author{Zhibo Feng}
\author{Zhuo Chen}
\author{Bruce Yuan}

\author[\dagger]{Harry Wang}
\author[\dagger]{Kris Chen} 
\affiliation{AgentAlpha Team}
\contribution[\dagger]{Corresponding author}

\abstract{
Autonomous scientific discovery with large language model (LLM)-based agents has recently made
substantial progress, demonstrating the ability to automate end-to-end research workflows.
However, existing systems largely rely on runtime-centric execution paradigms, repeatedly reading,
summarizing, and reasoning over large volumes of scientific literature online. This on-the-spot
computation strategy incurs high computational cost, suffers from context window limitations, and
often leads to brittle reasoning and hallucination. We propose Idea2Story, a pre-computation–driven framework for autonomous scientific discovery
that shifts literature understanding from online reasoning to offline knowledge construction.
Idea2Story continuously collects peer-reviewed papers together with their review feedback, extracts
core methodological units, composes reusable research patterns, and organizes them into a structured
methodological knowledge graph. At runtime, underspecified user research intents are aligned to
established research paradigms, enabling efficient retrieval and reuse of high-quality research
patterns instead of open-ended generation and trial-and-error. By grounding research planning and execution in a pre-built knowledge graph, Idea2Story alleviates
the context window bottleneck of LLMs and substantially reduces repeated runtime reasoning over
literature. We conduct qualitative analyses and preliminary empirical studies demonstrating that
Idea2Story can generate coherent, methodologically grounded, and novel research patterns, and can
produce several high-quality research demonstrations in an end-to-end setting. These results
suggest that offline knowledge construction provides a practical and scalable foundation for
reliable autonomous scientific discovery. Our codebase is publicly available at
\url{https://github.com/AgentAlphaAGI/Idea2Paper.git}.
}

\date{\today}

\begin{document}

\maketitle

\section{Introduction}

As research increasingly moves toward fully autonomous scientific discovery, large language model (LLM)-based agents have attracted growing attention for their ability to automate complex research workflows~\citep{chai2025scimaster, cornelio_combining_2023, wang2023scientific,xu_artificial_2021}. Recent systems ~\citep{lu2024aiscientist,yamada2025aiscientistv2,gottweis_towards_2025} demonstrate that LLM-based agents can autonomously execute an end-to-end research loop, including literature review, code generation, experiment execution, and manuscript drafting. These results suggest that automated scientific discovery is becoming practically feasible and that LLM-based agents are approaching a level of functional completeness required for autonomous research~\citep{jin_agentreview_2024,sahu_reviewertoo_2025,ajith2024litsearch,zhang_noveltybench_2025,zhang2026opennovelty}.

Despite this progress, existing systems remain constrained by a fundamental inefficiency in their execution paradigm, which limits their scalability and robustness in practice. In particular, most current research agents \citep{wang_openhands_2025, yang_swe-agent_2024, mitchener_kosmos_2025, luo2025llm4sr} rely on an \emph{on-the-spot computation} strategy, where nearly all information acquisition, reasoning, and synthesis are performed online at runtime. Under this paradigm, each new research attempt requires the agent to dynamically retrieve large volumes of scientific literature, read and summarize long and heterogeneous documents in real time, and explore a broad space of candidate methods and experimental designs through open-ended generation and trial-and-error. As a result, the cost of producing a single effective scientific discovery remains substantial. For example, a complete execution of the overall pipeline often requires several hours and, in some cases, up to 15 hours to progress from ideation to experimentation~\citep{lu2024aiscientist}. Similarly, in \citep{schmidgall_agent_2025}, literature review and experimental planning alone account for a significant portion of total inference time and place heavy demands on the language model’s ability to maintain coherent reasoning over long contexts. More importantly, this runtime-centric design repeatedly forces the model to re-process large volumes of unstructured and partially redundant information, even when much of the underlying scientific knowledge is already well established, thereby increasing computational overhead and exacerbating the risk of hallucination and reasoning errors \citep{wang2025repomaster, shin_mind_2025}.

To address the efficiency and reliability limitations of existing autonomous research agents, we propose Idea2Story, a scientific discovery framework that explicitly separates offline knowledge construction from online research generation, with the goal of reducing \emph{repeated reasoning over scientific literature} and alleviating the \emph{context window bottleneck} of large language models. Most current systems rely on runtime-centric execution, where agents repeatedly retrieve, read, summarize, and reason over large collections of highly overlapping papers for each new research attempt, resulting in substantial computational cost and prolonged execution time. Idea2Story mitigates this inefficiency by shifting literature understanding from online reasoning to an offline stage. In the offline phase, the system periodically collects recently accepted, peer-reviewed papers together with their full review feedback, extracts core methodological units and research patterns, and organizes these units and their observed composition relations into a continuously updated structured knowledge graph. This knowledge graph serves as a compact and reusable representation of established scientific methods and their empirical compatibility, replacing repeated processing of raw documents at runtime. Building on this offline knowledge infrastructure, Idea2Story performs online research generation by aligning underspecified user research intents with existing research paradigms encoded in the knowledge graph. Rather than relying on open-ended generation and trial-and-error, the system retrieves high-quality research patterns as structured compositions of method units, which act as stable methodological blueprints for downstream experimental design and execution. Guided by these validated research patterns, Idea2Story conducts feasibility-driven experimentation and ultimately generates a complete, submission-ready paper in an end-to-end manner.

% \begin{figure}[t]
%     \centering
%     \includegraphics[width=\linewidth]{picture/idea2story1-2_cropped.pdf}
%     \vspace{-5mm}
%     \caption{The offline knowledge graph construction process of Idea2Story. Starting from a curated pool of peer-reviewed papers, the system extracts core method units, composes them into research patterns, and organizes them into a structured knowledge graph.}
%     \label{fig:idea2story-pipeline}
% \end{figure}

% \begin{figure}[t]
%     \centering
%     \includegraphics[width=\linewidth]{picture/idea2story1-2.pdf}
%     \vspace{-5mm}
%     \caption{Overview of the proposed two-stage framework for automated scientific discovery. The upper panel illustrates offline knowledge graph construction, including paper pool curation and cleaning, method unit extraction, research pattern composition, and curated knowledge graph building with review feedback. The lower panel depicts online scientific discovery, where the system performs idea generation and paradigm alignment, feasibility search for experimental design via agent-based exploration, paper instantiation, and iterative review and refinement until final submission.}
%     \label{fig:idea2story-pipeline}
% \end{figure}

\begin{figure}[t]
    \centering
    \includegraphics[width=\linewidth]{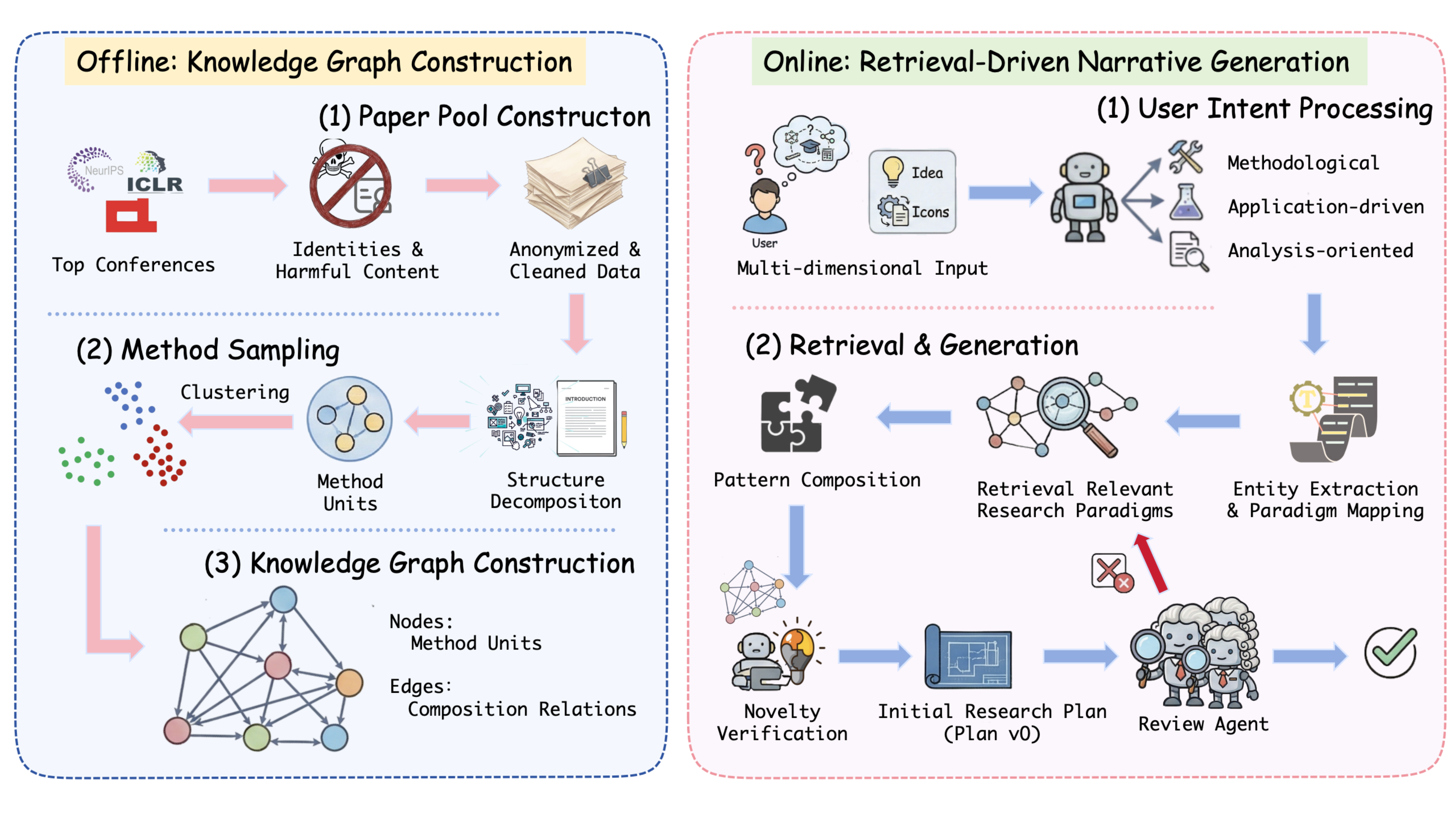}
    \vspace{-5mm}
    \caption{ Overview of the two-stage framework in Idea2Story. The offline stage constructs a structured knowledge graph by extracting and organizing reusable method units from a curated paper corpus. The online stage retrieves and composes research patterns from the knowledge graph to ground underspecified user intent into concrete and coherent research directions.}
    \label{fig:idea2story-pipeline}
\end{figure}

Our work makes the following contributions to autonomous scientific discovery :
(1) We introduce Idea2Story, a framework that formalizes autonomous research as a
\emph{pre-computation–driven} process, where scientific knowledge is extracted, structured, and
maintained in a continuously updated methodological knowledge graph, addressing the inefficiency and
unreliability of runtime-centric research agents. (2) We propose a knowledge-grounded planning and execution pipeline that alleviates the \emph{context window bottleneck} and reduces \emph{repeated runtime reasoning} over literature by converting paper reading into retrieval over a pre-built knowledge graph.   (3) We conduct preliminary empirical studies and comparative evaluations, demonstrating that Idea2Story can produce several high-quality research demos and establishing the practical feasibility of the proposed paradigm in an end-to-end setting.

% Building on this paradigm, Idea2Story formalizes the process of scientific discovery via a tightly integrated
% experimental pipeline. The system instantiates retrieved research patterns into candidate experimental
% paths and performs a feasibility-driven search across methodological and experimental design alternatives. Instead of
% relying on open-ended generation, this search is guided by explicit experimental validation, progressively
% filtering for designs that are both executable and empirically meaningful. During experimentation, specialized agents operate in a coordinated manner. An experimental design agent constructs comparable experimental protocols, including dataset selection, baseline methods, and evaluation metrics. An experimental execution agent implements and executes the experiments while monitoring correctness and stability. An experimental evaluation agent aggregates and analyzes the outcomes to determine whether a candidate path meets feasibility criteria. Through this validation-driven search process,
% Idea2Story efficiently discovers feasible experimental paths under constrained computational budgets.
% Upon identifying a feasible path, the system analyzes the resulting experimental evidence and automatically
% synthesizes the full manuscript in a section-by-section manner.

 % We design an end-to-end closed-loop system composed of specialized agents for experimental design, execution, paper writing, and review simulation, which enforces consistency across reasoning, experimentation, and manuscript generation while reusing validated intermediate results.

\section{Related Work}

\subsection{Autonomous Scientific Discovery}
%team2025tongyi
Recent advances in large language models (LLMs) have driven growing interest in autonomous scientific
discovery agents that aim to automate the full research lifecycle, from code generation to experimental
execution ~\citep{hu_controlled_2026, zhang2025evolving, lin_se-agent_2025}. Early systems such as \emph{The AI Scientist} (v1)~\citep{lu2024aiscientist} demonstrate the
viability of end-to-end automation but rely heavily on manually crafted code templates and largely
linear exploration workflows, which restrict discovery depth and adaptability. Later approaches, including
\emph{The AI Scientist-v2}~\citep{yamada2025aiscientistv2} and \emph{Kosmos}~\citep{mitchener_kosmos_2025}, reduce reliance on
explicit template through the incorporation of agentic tree search and experiment management agents, enabling iterative and multi-round exploration.

In research ideation, LLM-generated ideas are often perceived as highly novel during initial screening; however, prior studies \citep{si2024can} uncover a critical paradox whereby such ideas tend to underperform after implementation relative to human-generated ideas, indicating limited feasibility and practical
utility. As more ideas are generated, LLM outputs exhibit growing similarity, leading to diminished meaningful diversity. Similar limitations have also been observed in research evaluation and peer
review \citep{liang2024can, xu2025can,  thakkar_can_2025, zhang2026opennovelty}. Existing AI-based reviewers display systematic blind
spots: \cite{shin_mind_2025} shows that LLM reviewers place disproportionate
emphasis on technical correctness while undervaluing novelty, deviating from human
expert judgment, while \cite{sahu_reviewertoo_2025} demonstrates that AI reviewers
struggle to distinguish fine-grained acceptance categories and are susceptible to sycophancy, with
review scores increasing unreasonably after exposure to author rebuttals. Although recent approaches
such as AgentReview \citep{jin_agentreview_2024} seek to mitigate these deficiencies by simulating
diverse reviewer roles, automated evaluation systems remain less reliable than human experts in
identifying robust accept/reject decision boundaries.

\subsection{LLM-Driven Agents}

LLM-driven agents still struggle to interact effectively with complex real-world environments.
Despite their strong generative capabilities, many existing systems—such as OpenHands~\citep{wang_openhands_2025}
and SWE-Agent~\citep{yang_swe-agent_2024}—exhibit limited performance when applied to realistic
codebases. These limitations largely stem from insufficient reasoning over hierarchical dependencies
and structural constraints, as well as the inherent restrictions imposed by finite context windows.
As a result, LLM-driven agents achieve relatively low task completion rates on challenging benchmarks
such as \emph{MLE-bench}~\citep{chan_mlebench_2024} and \emph{SciCode}~\citep{tian_scicode_2024}.
RepoMaster~\citep{wang2025repomaster} further identifies inadequate modeling of codebase structure,
including function call graphs and module dependency graphs, as a key bottleneck for LLM-driven agents
operating in large and complex environments.

Beyond execution limitations, LLM-driven agents also exhibit notable deficiencies in scientific rigor
and evaluative judgment. When tasked with autonomous assessment, these agents are prone to hallucination and overconfidence. For instance, Agent Laboratory~\citep{schmidgall_agent_2025} reports that automated evaluations produced by LLM-driven agents substantially overestimate paper quality compared to human reviewers. Evaluations of \emph{Kosmos}~\citep{mitchener_kosmos_2025} further reveal a tendency to invent opaque quantitative metrics and to conflate statistical significance with scientific value, leading to weak interpretability of experimental conclusions. Moreover, long-horizon autonomous execution exacerbates these issues by introducing behavioral
drift~\citep{arike2025tech}, where LLM-driven agents gradually deviate from intended research trajectories or generate overly strong and insufficiently justified claims~\citep{lu2024aiscientist,
schmidgall2025agent, baek_researchagent_2025, hong_metagpt_2023, wu_autogen_2023,
lin_se-agent_2025, hu_controlled_2026}. This drift further undermines reliability and highlights the
need for stronger structural grounding and validation mechanisms in LLM-based autonomous research
systems.

\section{General Idea Generation}
\label{sec:stage1}

Idea2Story is designed to interact with users through high-level and often informal research ideas
that reflect human intuition rather than fully specified technical plans. The system transforms
such underspecified inputs into structured and academically grounded research directions through
a two-stage paradigm that separates offline knowledge construction from online research generation:

\begin{itemize}
    \item \textbf{Offline Knowledge Construction.}
    In the offline stage, Idea2Story builds a reusable methodological foundation from existing
    scientific literature. This includes curating a large-scale paper pool from peer-reviewed
    venues, extracting reusable method units that capture core methodological contributions, and
    organizing these units into a structured knowledge graph that encodes their semantic and
    compositional relations. The resulting knowledge graph serves as a persistent repository of
    methodological abstractions, decoupling literature understanding from runtime reasoning.

    \item \textbf{Online Research Generation.}
    In the online stage, Idea2Story grounds user-provided research ideas through retrieval and
    composition over the pre-built knowledge graph. Given an informal user idea, the system aligns
    the input with existing research paradigms, retrieves relevant research patterns, and composes
    compatible method units into concrete research directions. These instantiated patterns are
    further refined through a review-guided process that iteratively evaluates and revises them with
    respect to novelty, methodological soundness, and conceptual coherence. The refined research
    patterns then serve as structured blueprints for subsequent planning, feasibility-driven
    experimentation, and end-to-end paper generation.
\end{itemize}

\subsection{Offline Knowledge Construction}

The offline knowledge construction stage aims to distill reusable methodological structure from
existing scientific literature and to organize it in a form that can be efficiently accessed during
online research generation. Instead of performing document-level reasoning at runtime, Idea2Story
pre-computes a structured representation of prior work that captures both methodological
abstractions and their observed compatibility in accepted research. This stage consists of three
main components: (i) constructing a curated paper pool from peer-reviewed venues, (ii) extracting
core method units that represent reusable methodological contributions, and (iii) organizing these
units and their composition relations into a structured knowledge graph. Together, these components
form a persistent methodological memory that decouples literature understanding from downstream
idea grounding and research generation.

\subsubsection{Paper Pool Construction}

We construct a paper pool from accepted machine learning papers and their associated peer reviews
collected from top-tier conferences. Let $\mathcal{C} = \{\text{NeurIPS}, \text{ICLR}\}$ denote the
set of venues considered, and let $\mathcal{T}$ denote the most recent three-year time window.
The resulting paper pool is defined as
\[
\mathcal{P} = \{\, p \mid p \text{ is an accepted paper from } c \in \mathcal{C} \text{ during } \mathcal{T} \,\},
\]
which consists of approximately 5{,}000 papers from NeurIPS and 8{,}000 papers from ICLR. For each paper $p \in \mathcal{P}$, we retain the full textual content
\[
\mathbf{x}_p = (\text{title}_p, \text{abstract}_p, \text{body}_p),
\]
together with its associated review artifacts
\[
\mathbf{r}_p = \{\text{comments}, \text{ratings}, \text{confidence scores}, \text{meta-reviews}\}.
\]
This yields a temporally aligned corpus that jointly captures research contributions and evaluation
signals.

To protect privacy, we apply an anonymization function $\mathcal{A}(\cdot)$ that removes all
author- and reviewer-identifying information, including names, affiliations, email addresses, and
explicit identity references. In addition, we apply a safety filtering function
$\mathcal{F}(\cdot)$ to review content to remove toxic or abusive language and personal attacks.
The final stored representation of each paper is given by
\[
\tilde{p} = \mathcal{F}(\mathcal{A}(p)),
\]
resulting in a de-identified paper pool
\[
\tilde{\mathcal{P}} = \{\, \tilde{p} \mid p \in \mathcal{P} \,\},
\]
which preserves technical content and review feedback while minimizing exposure to private or
harmful information.

\subsubsection{Method Unit Extraction}

Based on the de-identified paper pool $\tilde{\mathcal{P}}$, we define an automated extraction
procedure that identifies the core methodological contributions of each paper in a structured and
reusable form. Formally, we model method unit extraction as a mapping
\[
\mathcal{E} : \tilde{p} \rightarrow \mathcal{U}_p = \{u_{p}^{(1)}, \dots, u_{p}^{(K_p)}\},
\]
where $\tilde{p} \in \tilde{\mathcal{P}}$ denotes a single paper and $\mathcal{U}_p$ is a small set
of method units that capture its essential technical ideas.

As illustrated in Figure~2, the extraction procedure leverages the standardized structure of
academic papers and analyzes different sections to collect complementary methodological signals.
Let $\mathbf{x}_p = (\text{intro}_p, \text{method}_p, \text{exp}_p)$ denote the partition of a paper
into its introduction, method, and experiments sections. The introduction is used to identify the
high-level research motivation and the precise problem formulation, the method section provides
signals about core technical mechanisms such as modeling assumptions, learning objectives, model
architectures, and optimization strategies, and the experiments section reflects how these
mechanisms are instantiated and evaluated in practice. By jointly aggregating information from
these sections, the extractor isolates method units that correspond to the primary algorithmic or
modeling contributions of the paper, rather than surface-level experimental details.

We define a method unit $u \in \mathcal{U}_p$ as a self-contained description of how a research
problem is formulated or solved, abstracted away from specific implementation choices and
experimental configurations. Elements that primarily involve dataset selection, hyperparameter
tuning, or engineering-level optimizations are excluded unless they induce substantive changes to
the problem formulation, model structure, or learning objective. In practice, most papers yield one
or a small number of method units. Each extracted unit is further normalized into structured
methodological attributes, including \emph{atomic meta-methods}, which correspond to indivisible
methodological elements, and \emph{composition-level patterns}, which describe how multiple method
units are combined within a single paper.

% After extracting method units for all papers, we compute vector embeddings for each paper based on
% its associated method units. Let $\mathbf{z}_p$ denote the resulting embedding of paper $p$.
% We apply UMAP to project the high-dimensional representations $\{\mathbf{z}_p\}$ into a
% lower-dimensional space that preserves local semantic structure, and then perform density-based
% clustering using DBSCAN on the reduced representations. This procedure groups semantically related
% papers into coherent research patterns, which serve as higher-level abstractions for subsequent
% retrieval and composition.

After extracting method units for all papers, we represent each paper $p \in \tilde{\mathcal{P}}$
by a vector embedding derived from its associated method units. Formally, let
\[
\mathbf{z}_p = g(\mathcal{U}_p),
\]
where $\mathcal{U}_p$ denotes the set of extracted method units for paper $p$ and
$g(\cdot)$ is an embedding function that maps a set of method units to a fixed-dimensional
representation.

To induce higher-level research patterns, we first apply a nonlinear dimensionality reduction
operator
\[
\mathbf{y}_p = \mathrm{UMAP}(\mathbf{z}_p),
\]
which projects the high-dimensional embeddings into a lower-dimensional space while preserving
local semantic neighborhoods. We then perform density-based clustering on the reduced
representations using DBSCAN, yielding a partition
\[
\mathcal{C} = \{ C_1, \dots, C_M \},
\]
where each cluster $C_m \subset \tilde{\mathcal{P}}$ corresponds to a coherent research pattern.

These induced clusters serve as higher-level abstractions over individual papers, capturing
recurring methodological structures that are reused across the literature. The resulting research
patterns form the basis for subsequent retrieval and composition.

\begin{figure}[t]
    \centering
    \includegraphics[width=\linewidth]{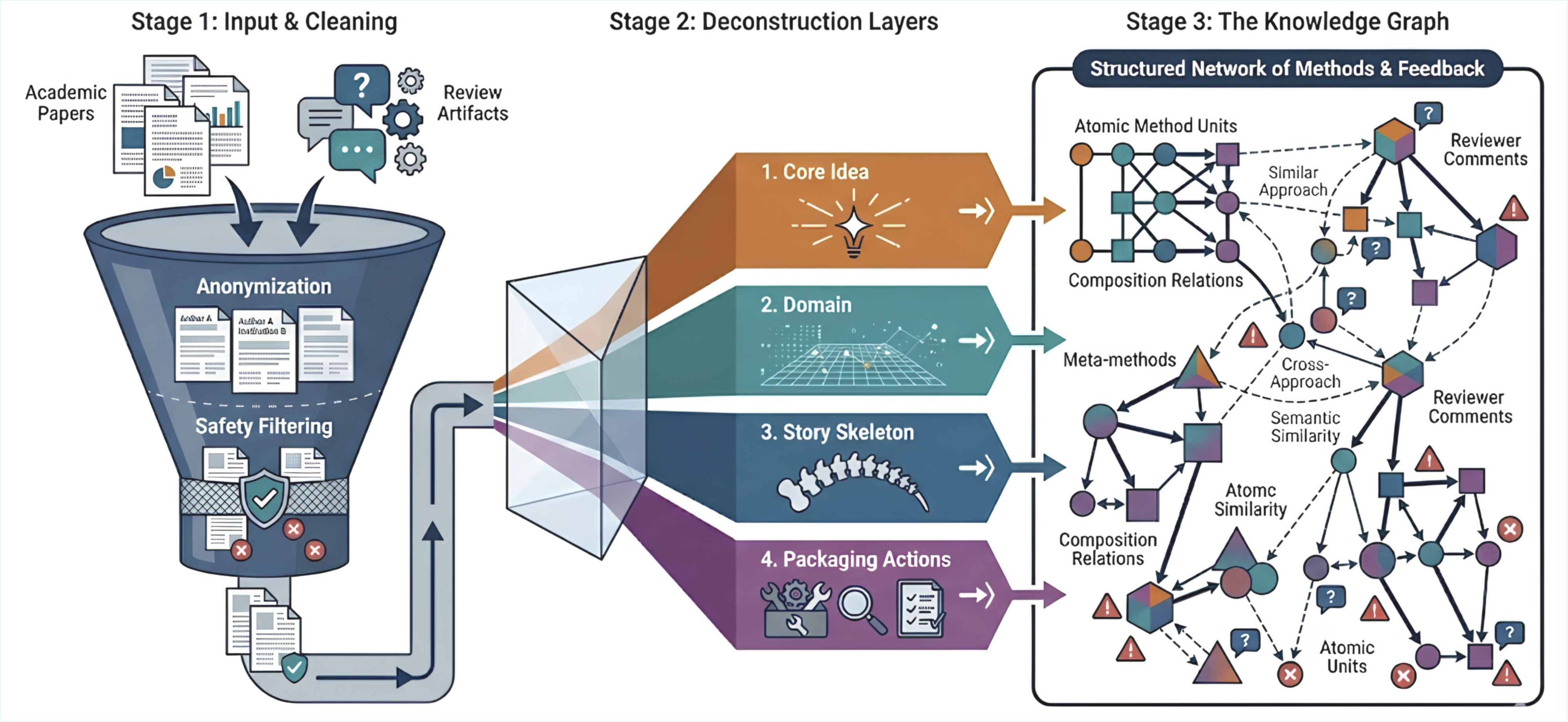}
    % \vspace{-8mm}
    \caption{ Offline knowledge graph construction in Idea2Story. Academic papers and their associated review artifacts are first anonymized and safety-filtered, then deconstructed into layered methodological representations. These layers capture complementary aspects of a paper, including its core research idea, domain context, high-level story skeleton, and packaging actions. The extracted elements are normalized into atomic method units and meta-methods, which are connected through composition and similarity relations. Reviewer feedback is incorporated as additional signals to refine relations and validate abstractions.   }
    \label{fig:phase0}
\end{figure}

\subsubsection{Knowledge Graph Construction}

Building on the extracted method units, we organize reusable methodological components into a
structured knowledge graph that supports systematic method discovery and composition. While
individual method units capture isolated algorithmic or modeling ideas, effective research methods
in practice typically arise from structured combinations of multiple method units. The knowledge
graph provides a unified representation that explicitly encodes canonicalized method units,
meta-methods, and their empirically observed composition relations in prior work.

Formally, we define the knowledge graph as a directed graph
\[
\mathcal{G} = (\mathcal{V}, \mathcal{E}),
\]
where each node $v \in \mathcal{V}$ corresponds to a canonicalized method unit or a meta-method.
Canonicalization groups semantically similar method units across the corpus into shared
meta-method abstractions, reducing surface-level variation while preserving core methodological
intent. As a result, nodes in the graph represent atomic or minimally indivisible methodological
elements that are reused across papers.

Edges in the graph encode composition relations between method units. For a given paper
$p \in \tilde{\mathcal{P}}$ with extracted method unit set $\mathcal{U}_p$, we add directed edges
between pairs of method units $(u_i, u_j) \in \mathcal{U}_p \times \mathcal{U}_p$ to indicate that
they are jointly instantiated as part of the same methodological pipeline. These edges capture
empirical evidence of method compatibility observed in prior work, reflecting how different
method units are combined in practice rather than hypothetical or manually specified relations.

Aggregating composition relations across the full corpus yields a graph structure that encodes both
methodological abstraction and empirical compatibility. In particular, the graph captures two
complementary levels of structure: (i) reusable methodological elements represented as
canonicalized method units and meta-methods, and (ii) composition constraints induced from
co-occurrence statistics in accepted papers. This separation allows Idea2Story to reason about
methods at a higher level of abstraction than individual papers, while remaining grounded in
observed research practice.

% To support systematic discovery and composition of reusable methodological ideas, we organize the
% extracted method units into a structured knowledge graph. In this graph, each node corresponds to a
% method unit (or a meta-method), representing an atomic methodological component identified from the
% literature.

% For a given paper, the extraction agent typically identifies multiple method units, which together
% form a complete methodological pipeline. We explicitly record these intra-paper composition
% relations by connecting method units originating from the same paper, capturing how different
% components are jointly instantiated in practice. Across the full corpus, method units are further
% grouped based on semantic similarity to normalize surface-level variations and identify shared
% meta-method abstractions. The resulting knowledge graph is constructed by combining these
% canonicalized nodes with observed composition relations.

% By representing method units and their compositional structure in a unified graph, Idea2Story
% enables method discovery and assembly to be treated as a graph-based retrieval and composition
% process. Given a target research objective, the system retrieves relevant subgraphs and composes
% compatible method units into coherent research patterns. These research patterns serve as high-level
% methodological blueprints that guide downstream experimental design and provide structured support
% for end-to-end paper generation.

\subsection{Online Research Generation.}

Given a target research objective, Idea2Story treats method discovery as a graph-based retrieval and
composition problem over $\mathcal{G}$. The system retrieves relevant subgraphs and composes
compatible method units by following connectivity constraints in the graph, producing candidate
research patterns that correspond to structured combinations of method units. These research
patterns serve as high-level methodological blueprints that bridge abstract research intent and
concrete experimental design, enabling downstream planning, feasibility analysis, and end-to-end
paper generation.

\subsubsection{Research Pattern Retrieval}

Given a user-provided research idea expressed in natural language, we formulate research pattern
identification as a structured retrieval problem over the knowledge graph $\mathcal{G}$. Let
$q$ denote the input research idea, and let $\mathcal{C} = \{C_1, \dots, C_M\}$ denote the set of
research patterns induced from the paper corpus. The goal is to rank patterns in $\mathcal{C}$
according to their relevance to $q$.

Rather than relying on a single similarity metric, Idea2Story adopts a multi-view retrieval
formulation that aggregates complementary signals from different semantic abstractions. Formally,
for each research pattern $C_m$, we compute a relevance score
\[
s(C_m \mid q) = \sum_{v \in \mathcal{V}} \lambda_v \, s_v(C_m \mid q),
\]
where $\mathcal{V} = \{\text{idea}, \text{domain}, \text{paper}\}$ indexes the retrieval views,
$s_v(\cdot)$ denotes a view-specific scoring function, and $\lambda_v$ are fixed weighting
coefficients that balance the contribution of different views.

\paragraph{Idea-level retrieval.}
At the idea level, the system retrieves previously observed research ideas that are semantically
similar to the input query $q$. Let $\mathcal{I}$ denote the set of stored research ideas extracted
from the corpus, and let $\mathrm{sim}_{\text{idea}}(q, i)$ denote a semantic similarity function
between $q$ and an idea $i \in \mathcal{I}$. The idea-level score of a research pattern $C_m$ is
computed by aggregating the similarity scores of ideas associated with the pattern:
\[
s_{\text{idea}}(C_m \mid q) = \max_{i \in \mathcal{I}(C_m)} \mathrm{sim}_{\text{idea}}(q, i),
\]
where $\mathcal{I}(C_m)$ denotes the set of ideas linked to pattern $C_m$.

\paragraph{Domain-level retrieval.}
At the domain level, the system interprets the input idea $q$ in terms of its underlying research
domains and methodological themes. Let $\mathcal{D}$ denote the set of research domains, and let
$\mathrm{sim}_{\text{domain}}(q, d)$ measure the relevance between $q$ and domain $d \in \mathcal{D}$.
The domain-level score of pattern $C_m$ is computed as
\[
s_{\text{domain}}(C_m \mid q) = \sum_{d \in \mathcal{D}(C_m)} \mathrm{sim}_{\text{domain}}(q, d)
\, w(d, C_m),
\]
where $\mathcal{D}(C_m)$ denotes the domains associated with pattern $C_m$, and $w(d, C_m)$ captures
empirical effectiveness signals derived from the knowledge graph.

\paragraph{Paper-level retrieval.}
At the paper level, the system retrieves papers whose technical content is semantically aligned
with the input idea. Let $\mathcal{P}(C_m)$ denote the set of papers instantiating pattern $C_m$.
The paper-level score is computed as
\[
s_{\text{paper}}(C_m \mid q) =
\max_{p \in \mathcal{P}(C_m)} \mathrm{sim}_{\text{paper}}(q, p) \cdot \alpha(p),
\]
where $\mathrm{sim}_{\text{paper}}(q, p)$ measures semantic similarity between $q$ and paper $p$,
and $\alpha(p)$ denotes a quality-related weight derived from peer review metadata.

The final ranked list of research patterns is obtained by ordering patterns according to their
aggregated multi-view relevance scores. Formally, we define
\[
\mathcal{C}^*(q) =
\operatorname{Rank}_{C_m \in \mathcal{C}}
\left(
\sum_{v \in \{\text{idea},\text{domain},\text{paper}\}}
\lambda_v \, s_v(C_m \mid q)
\right),
\]
where patterns are sorted in descending order of the aggregated score.

\subsubsection{Review-Guided Refinement}

After candidate research patterns are retrieved, Idea2Story refines them using an explicit
LLM-based review loop. In each iteration, a large language model is prompted to act as a reviewer
and evaluate the current research pattern along several predefined criteria, including technical
soundness, novelty with respect to existing literature, and overall clarity of the problem–method
alignment. The reviewer produces both scalar judgments and concrete revision suggestions.

The system then uses this feedback to update the research pattern in a targeted manner. When the
review indicates insufficient novelty, the system modifies the pattern by recombining compatible
method units or introducing alternative realizations within the same pattern family. When the
review identifies issues in feasibility or ambiguity in formulation, the system revises the problem
definition or method structure to improve consistency and executability. Each revised pattern is
re-submitted to the same review process, forming an explicit generate–review–revise loop.

To prevent uncontrolled drift, only revisions that improve the reviewer scores are retained;
otherwise, the system rolls back to the previous version. This process repeats until the reviewer
judges the pattern to be sufficiently novel, coherent, and technically plausible, or until further
iterations no longer yield improvement. The output of this stage is a refined research pattern that
has been iteratively vetted by an LLM-based reviewer and is suitable for downstream validation and
paper generation.

\section{Experiments and Analysis}

We evaluate Idea2Story through a set of experiments focusing on its ability to extract reusable
methodological structure and to generate high-quality research patterns from ambiguous user input.
Our experiments are conducted on a corpus of accepted papers from ICLR and NeurIPS over the past
three years, including approximately 13K papers and their associated peer reviews, which serves as
the foundation for all subsequent analyses. Based on this corpus, we first analyze the properties of the extracted method units to assess whether Idea2Story captures meaningful and reusable methodological abstractions. We then present qualitative demonstrations of research patterns instantiated as structured research stories, illustrating how the system transforms vague research intent into coherent and methodologically grounded research directions.

\begin{figure}[t]
\centering
\begin{tcolorbox}[
  colback=gray!5,
  colframe=gray!40,
  title={Case 1: Method Unit Extraction Demo},
  width=1\linewidth,
  boxrule=0.6pt,
  sharp corners,
  breakable
]
\textbf{Paper Title:}  
Learning Dynamics of LLM Finetuning

\vspace{0.5em}
\textbf{Base Problem:}  
Understanding how specific training examples influence model predictions during finetuning is challenging, particularly in large language models.

\vspace{0.5em}
\textbf{Solution Pattern:}  
Develop a framework to analyze step-wise influence accumulation among potential responses during finetuning, providing insights into phenomena like hallucination and the squeezing effect in off-policy direct preference optimization.

\vspace{0.5em}
\textbf{Story:}  
Reframe the understanding of LLM finetuning through the lens of learning dynamics, offering a unified interpretation of training behaviors and inspiring methods to enhance model alignment and performance.

\vspace{0.5em}
\textbf{Application:}  
Improving alignment in large language models, enhancing finetuning strategies for better model performance, diagnosing and mitigating hallucination in AI systems.
\end{tcolorbox}

\vspace{-3mm}
\caption{An example of a method unit extracted from an accepted paper, illustrating the separation of the base problem, solution pattern, and higher-level research story.}
\label{fig:method-unit-demo}
\end{figure}

\subsection{Implementation Details}

To further assess the effectiveness of Idea2Story in practical research ideation settings, we
conduct additional qualitative experiments on a small set of representative cases. Specifically,
we evaluate three user-provided research ideas curated by an external collaborator. For each case,
Idea2Story generates research patterns using the GLM-4.7~\citep{zeng2025glm} model as the underlying language backbone. As a baseline, we compare against direct LLM generation, where the same model is prompted to produce a complete research story without explicit pattern modeling or retrieval.

% To reduce evaluation bias, the generated research stories from both approaches are subsequently
% assessed by an independent large language model (Gemini 3 Pro), which is not involved in either
% generation process. The evaluator is instructed to compare the outputs in terms of novelty,
% methodological substance, and overall research quality, without access to the generation method
% used.

% \begin{tcolorbox}[
%   colback=gray!5,
%   colframe=gray!40,
%   title={Case 1 : Method Unit Extraction Demo}
% ]
% \textbf{Paper Title : }  
% Learning Dynamics of LLM Finetuning

% \vspace{0.5em}
% \textbf{Base Problem : }  
% Understanding how specific training examples influence model predictions during finetuning is challenging, particularly in large language models.

% \vspace{0.5em}
% \textbf{Solution Pattern : }  
% Develop a framework to analyze step-wise influence accumulation among potential responses during finetuning, providing insights into phenomena like hallucination and the squeezing effect in off-policy direct preference optimization.

% \vspace{0.5em}
% \textbf{Story : }  
% Reframe the understanding of LLM finetuning through the lens of learning dynamics, offering a unified interpretation of training behaviors and inspiring methods to enhance model alignment and performance.

% \vspace{0.5em}
% \textbf{Application : } 
% Improving alignment in large language models, enhancing finetuning strategies for better model performance, diagnosing and mitigating hallucination in AI systems.
% \end{tcolorbox}

\subsection{Case Study: Method Unit Extraction}

We present a representative case study to illustrate the behavior of the proposed method unit
extraction agent. Case 1 shows an example extracted from an accepted paper, where the system decomposes the full paper into a structured set of methodological elements.

As shown in the example, the extracted method unit explicitly separates the underlying research
problem, the core solution pattern, and the resulting research story. The \textit{Base Problem} describes the core challenge addressed by the paper, namely understanding how individual training examples influence model behavior during finetuning, without depending on specific datasets or implementation details. The \textit{Solution Pattern} summarizes the central methodological idea as
an analysis framework for step-wise influence accumulation, highlighting the key mechanism without
binding it to a particular optimization setup or experimental configuration. Importantly, the extracted \textit{Story} reframes the technical contribution at a higher level of
abstraction, connecting learning dynamics to broader phenomena such as hallucination and alignment
in large language models. This abstraction reflects how the method unit goes beyond algorithmic
details to capture the conceptual contribution of the paper. Finally, the \textit{Application}
field grounds the method unit by indicating downstream research and system-level implications,
without enumerating task-specific benchmarks.

This example demonstrates that the extraction agent isolates reusable methodological structure while
filtering out implementation-level details. By representing the paper as a coherent method unit
rather than a collection of experimental components, Idea2Story enables subsequent reuse,
comparison, and composition of methodological ideas across papers.

\subsection{Knowledge Graph Analysis}

\begin{figure}[h]
\centering
\begin{minipage}[t]{0.48\linewidth}
\vspace{0pt}
We analyze the structure of the constructed knowledge graph to understand how extracted method
units are distributed across papers and research domains. As illustrated in Figure~2, the graph
exhibits a clear hub-and-spoke structure, where a small number of high-frequency domains connect
to a large number of papers and research patterns. This reflects the uneven distribution of
research activity across domains, while also highlighting domains that function as central hubs
for methodological reuse. Importantly, many research patterns are observed to connect multiple
domains simultaneously, indicating that the extracted method units often capture methodological
abstractions that generalize beyond a single application area. In contrast, paper-level nodes are typically associated with a single domain, whereas pattern-level nodes frequently act as bridges between otherwise weakly connected domains. This structural separation suggests that the knowledge  graph encodes two distinct levels of organization—instance-level

\end{minipage}
\hfill
\begin{minipage}[t]{0.48\linewidth}
\vspace{0pt}
\centering
\includegraphics[width=\linewidth]{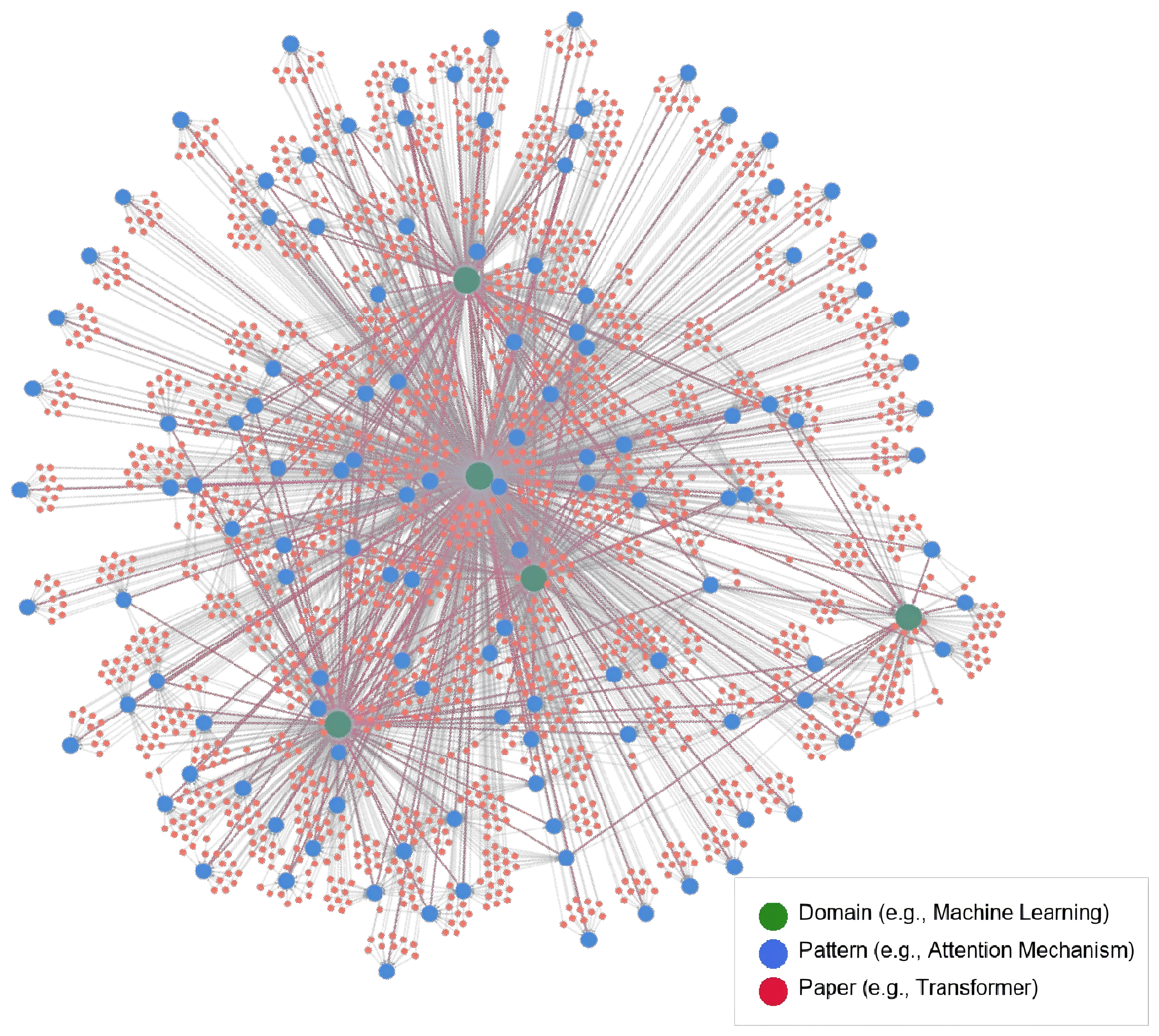}
\caption{Visualization of the knowledge graph substructure induced by high-frequency research
domains.}
\label{fig:knowledge_graph}
\end{minipage}
\end{figure}

 research artifacts and reusable methodological abstractions—enabling Idea2Story to retrieve and compose research patterns at a higher level of abstraction rather than relying on domain-specific or paper-specific similarity alone.

\begin{table*}[t!]
\centering
\small
\renewcommand{\arraystretch}{1.2}
\begin{tabular}{p{3.0cm} p{6.6cm} p{6.6cm}}
\toprule
\textbf{Aspect} &
\textbf{Idea2Story Generated (IntentDiff)} &
\textbf{LLM Direct Generated (EcoIntent)} \\
\midrule

Title &
\textbf{IntentDiff}: Reframing E-commerce Intent Classification via Structural Evolution and Context-Aware Diffusion &
\textbf{EcoIntent}: A Context-Aware Multi-Granularity Agent for E-commerce Intent Understanding via Hierarchical Contrastive Learning \\

\midrule

Abstract Focus &
Reinterprets intent classification as a structural evolution process rather than static text classification. The approach leverages a diffusion-based framework to iteratively refine noisy query representations into precise intent labels, integrates product graph embeddings to ground predictions in e-commerce context, and introduces a discrete, context-aware tokenizer to handle long-tail domain vocabulary. &
Targets improved intent classification performance by integrating heterogeneous behavioral context and hierarchical product knowledge. A dual-stream architecture aligns semantic representations with user interaction history, and hierarchical contrastive learning enforces consistency across fine- and coarse-grained intent categories. \\

\midrule

Problem Definition &
Reframes e-commerce intent classification from static text prediction to dynamic structural reasoning. User queries are short, ambiguous, and heavily dependent on implicit catalog structure, which fixed-label classification fails to capture. Intent understanding is modeled as an evolving process under structural constraints. &
Formulates intent understanding as a conventional multi-class classification problem, where the input is a query augmented with session context and the output is an intent label from a predefined set. The main challenge is semantic sparsity caused by short and ambiguous queries. \\

\midrule

Core Research Gap &
Existing intent classification methods treat queries in isolation and ignore domain-specific structural priors in e-commerce. They fail to exploit rich relationships between products and attributes, and standard vocabularies struggle with long-tail, domain-specific terminology. No prior work unifies diffusion-based refinement with structural graph embeddings for intent disambiguation. &
Prior work suffers from (1) context isolation, where behavioral signals such as clicks are underutilized, and (2) a flat-label assumption that ignores the hierarchical nature of e-commerce taxonomies, leading to inconsistent predictions for fine-grained, long-tail intents. \\

\midrule

Method Skeleton &
A diffusion-based classifier that iteratively denoises intent representations; a context-aware discrete tokenizer based on a VQ-VAE variant to encode diverse e-commerce queries; and integration of pretrained product graph embeddings as structural priors during the denoising process. &
A dual-stream discriminative architecture consisting of a BERT-based text encoder, a lightweight GNN for aggregating behavioral interaction graphs, and a prediction head trained with hierarchical contrastive learning; parameter-efficient adaptation via LoRA. \\

\midrule

Innovation Claims &
(1) Reformulates intent classification as a diffusion-based dynamic refinement process;  
(2) Introduces discrete, context-aware intent tokenization to better handle long-tail domain vocabulary;  
(3) Enhances intent reasoning by incorporating product graph structural embeddings. &
(1) Contextualized intent modeling via joint reasoning over text and behavioral graphs;  
(2) Hierarchical contrastive learning leveraging product taxonomies;  
(3) Parameter-efficient system design achieving strong performance at reduced computational cost. \\

\bottomrule
\end{tabular}
\vspace{-2mm}
\caption{
Comparison of research patterns generated by Idea2Story and a direct LLM baseline,
both starting from the same underspecified user input:
\emph{``I want to build an e-commerce agent that can better understand user intent.''}
The table contrasts how different generation mechanisms transform the same vague research intent
into concrete research patterns.
}
\label{tab:Idea2Story_vs_llm_english}
\end{table*}

\subsection{Qualitative Comparison of Generated Research Patterns}

We further compare the quality of research patterns generated by Idea2Story and a direct LLM
baseline. Both systems start from the same underspecified user input and produce structured
research proposals, enabling a controlled comparison of how different generation mechanisms
transform vague research intent into concrete research patterns.

Table~1 presents a side-by-side comparison of representative outputs along multiple dimensions,
including problem formulation, methodological structure, and innovation claims. Rather than
evaluating surface-level writing quality, the comparison focuses on the resulting research
patterns as methodological blueprints—i.e., how the generated ideas frame the research problem,
identify gaps in prior work, and organize methodological components into a coherent approach. As shown in the table, Idea2Story tends to induce higher-level problem reformulation, transforming
intent understanding from a fixed classification task into a dynamic structural reasoning process.
The resulting research pattern emphasizes generative refinement, structural priors, and evolving
representations. In contrast, the direct LLM baseline largely operates within a conventional task
formulation, proposing a stronger system through the integration of additional components such as
context modeling and hierarchical objectives.

% These differences suggest that Idea2Story contributes primarily at the level of research pattern
% abstraction and problem framing, whereas direct LLM generation focuses on optimizing solutions
% within predefined problem settings. Importantly, both outputs are reasonable and technically
% plausible; however, they reflect distinct modes of research ideation, highlighting the role of
% explicit pattern modeling in shaping the quality of generated research directions.

To reduce evaluation bias, the generated research stories from both approaches are subsequently
assessed by an independent large language model (Gemini 3 Pro)~\citep{team2025gemma}, which is not involved in either generation process. The evaluator is instructed to compare the outputs in terms of novelty, methodological substance, and overall research quality, without access to the generation method
used. Across all evaluated cases, the externally evaluated results consistently favor the outputs
generated by Idea2Story. In particular, the research stories produced by direct LLM generation tend
to remain at a high level of abstraction, with less concrete methodological grounding and reliance
on relatively standard techniques. In contrast, Idea2Story-generated research patterns exhibit
clearer problem framing, more specific methodological structures, and stronger signals of novelty.

\section{Future Work}

While Idea2Story focuses on grounding vague research intent into structured and high-quality research patterns, an important direction for future work is to extend this framework toward a fully closed-loop research generation pipeline. A promising extension is the integration of experiment-driven agents that can instantiate, validate, and iteratively refine generated research patterns through empirical feedback, including automated experimental design, dataset selection, and preliminary execution. Experimental outcomes can then serve as additional signals to refine the instantiated research stories, forming a feedback loop between method design and empirical validation. Beyond experimentation, future work may further explore how refined research patterns can be systematically translated into complete paper drafts, covering method descriptions, experimental results, and discussion sections. By grounding paper generation in empirically validated research patterns, such a system could move beyond surface-level text generation and provide more faithful, end-to-end support for executable and publishable scientific discovery.

\section{Conclusion}

We presented Idea2Story, a pre-computation–driven framework for autonomous scientific discovery that shifts literature understanding from runtime reasoning to offline knowledge structuring. By explicitly extracting reusable method units and organizing them into a continuously updated knowledge graph, Idea2Story enables research agents to reason over stable research patterns rather than repeatedly processing raw papers. Our qualitative analyses and comparative studies show that this design leads to research patterns with clearer problem reformulation, stronger methodological structure, and higher conceptual novelty than direct LLM generation. These results highlight the importance of explicit pattern modeling as a foundation for scalable and reliable autonomous research. Looking ahead, integrating Idea2Story with experimental agents to close the loop from abstract research patterns to validated empirical results represents a promising direction toward fully autonomous and trustworthy scientific discovery.

\clearpage
\newpage
\bibliographystyle{assets/plainnat}
\bibliography{paper}

@article{wang2023scientific,
  title={Scientific discovery in the age of artificial intelligence},
  author={Wang, Hanchen and Fu, Tianfan and Du, Yuanqi and Gao, Wenhao and Huang, Kexin and Liu, Ziming and Chandak, Payal and Liu, Shengchao and Van Katwyk, Peter and Deac, Andreea and others},
  journal={Nature},
  volume={620},
  number={7972},
  pages={47--60},
  year={2023},
  publisher={Nature Publishing Group UK London}
}

@misc{yamada2025aiscientistv2,
      title={The AI Scientist-v2: Workshop-Level Automated Scientific Discovery via Agentic Tree Search}, 
      author={Yutaro Yamada and Robert Tjarko Lange and Cong Lu and Shengran Hu and Chris Lu and Jakob Foerster and Jeff Clune and David Ha},
      year={2025},
      eprint={2504.08066},
      archivePrefix={arXiv},
      primaryClass={cs.AI},
      url={https://arxiv.org/abs/2504.08066}, 
}

@misc{gottweis_towards_2025,
	title = {Towards an {AI} co-scientist},
	journal={arXiv preprint arXiv:2502.18864},
	doi = {10.48550/arXiv.2502.18864},
	urldate = {2025-10-15},
	publisher = {arXiv},
	author={Juraj Gottweis and Wei-Hung Weng and Alexander Daryin and Tao Tu and Anil Palepu and Petar Sirkovic and Artiom Myaskovsky and Felix Weissenberger and Keran Rong and Ryutaro Tanno and others},
	month = feb,
	year = {2025},
	note = {arXiv:2502.18864 [cs]},
}

@article{ajith2024litsearch,
  title={Litsearch: A retrieval benchmark for scientific literature search},
  author={Ajith, Anirudh and Xia, Mengzhou and Chevalier, Alexis and Goyal, Tanya and Chen, Danqi and Gao, Tianyu},
  journal={arXiv preprint arXiv:2407.18940},
  year={2024}
}

@article{cornelio_combining_2023,
  title={Combining data and theory for derivable scientific discovery with {AI}-Descartes},
  author={Cornelio, Cristina and Dash, Sanjeeb and Austel, Vernon and Josephson, Tyler R and Goncalves, Joao and Clarkson, Kenneth L and Megiddo, Nimrod and El Khadir, Bachir and Horesh, Lior},
  journal={Nature Communications},
  volume={14},
  number={1},
  pages={1777},
  year={2023},
  publisher={Nature Publishing Group}
}

@article{xu_artificial_2021,
  title={Artificial intelligence: A powerful paradigm for scientific research},
  author={Xu, Yanjie and Liu, Xin and Cao, X and Huang, C and Liu, E and Qian, S and Liu, X and Wu, Y and Dong, F and Qiu, CW and others},
  journal={{The Innovation}},
  volume={2},
  number={4},
  pages={100179},
  year={2021},
  publisher={Cell Press}
}

@article{lu2024aiscientist,
  title={The {AI} {S}cientist: Towards Fully Automated Open-Ended Scientific Discovery},
  author={Lu, Chris and Lu, Cong and Lange, Robert Tjarko and Foerster, Jakob and Clune, Jeff and Ha, David},
  journal={arXiv preprint arXiv:2408.06292},
  year={2024}
}

@article{schmidgall2025agent,
  title={Agent laboratory: Using llm agents as research assistants},
  author={Schmidgall, Samuel and Su, Yusheng and Wang, Ze and Sun, Ximeng and Wu, Jialian and Yu, Xiaodong and Liu, Jiang and Liu, Zicheng and Barsoum, Emad},
  journal={arXiv preprint arXiv:2501.04227},
  year={2025}
}

@article{liang2024can,
  title={Can Large Language Models Provide Useful Feedback on Research Papers? A Large-Scale Empirical Analysis},
  author={Liang, Weixin and Zhang, Yuhui and Cao, Hancheng and Wang, Binglu and Ding, Daisy Yi and Yang, Xinyu and Vodrahalli, Kailas and He, Siyu and Smith, Daniel Scott and Yin, Yian and others},
  journal={{NEJM} {AI}},
  volume={1},
  number={8},
  pages={AIoa2400196},
  year={2024},
  publisher={Massachusetts Medical Society}
}

@article{thakkar_can_2025,
  title={Can LLM feedback enhance review quality? A randomized study of 20K reviews at ICLR 2025},
  author={Thakkar, Naitian and Xu, Yilun and Varma, Shikhar and Wu, Ke and Wang, Zhaofeng and Song, Dawn and Xu, Huazhe and Darrell, Trevor and Wang, Shanghang and Gonzalez, Joseph E},
  journal={arXiv preprint arXiv:2504.09737},
  year={2025}
}

@article{wang_openhands_2025,
      title={{OpenHands}: An Open Platform for {AI} Software Developers as Generalist Agents}, 
      author={Xingyao Wang and Bowei Yang and Yiqiao Jin and Jiaqi Li and Yijia Xiao and Wenghua Lin and Xiaotian Cheng and Ruicheng Zheng and Huieu Le and Maosong Cao and others},
      journal={arXiv preprint arXiv:2407.16741},
      year={2025}
}

@article{yang_swe-agent_2024,
      title={{SWE}-agent: Agent-Computer Interfaces Enable Automated Software Engineering}, 
      author={John Yang and Carlos E. Jimenez and Alexander Wettig and Kilian Lieret and Shunyu Yao and Karthik Narasimhan and Ofir Press},
      journal={arXiv preprint arXiv:2405.15793},
      year={2024}
}

@article{schmidgall_agent_2025,
      title={{Agent Laboratory}: Using {LLM} Agents as Research Assistants}, 
      author={Samuel Schmidgall and Yusheng Su and Ze Wang and Ximeng Sun and Jialian Wu and Xiaodong Yu and Jiang Liu and Zicheng Liu and Emad Barsoum},
      journal={arXiv preprint arXiv:2501.04227},
      year={2025}
}

@article{mitchener_kosmos_2025,
      title={{Kosmos}: An {AI} Scientist for Autonomous Discovery}, 
      author={Ludovico Mitchener and Angela Yiu and Benjamin Chang and Mathieu Bourdenx and Tyler Nadolski and Arvis Sulovari and Eric C. Landsness and Daniel L. Barabasi and Siddharth Narayanan and Nicky Evans and Shriya Reddy and others},
      journal={arXiv preprint arXiv:2511.02824},
      year={2025}
}

@article{shin_mind_2025,
      title={Mind the Blind Spots: A Focus-Level Evaluation Framework for {LLM} Reviews}, 
      author={Hyungyu Shin and Jihoon Kim and Hwaran Lee and Kyohoon Jin and Seung-won Hwang},
      journal={arXiv preprint arXiv:2502.17086},
      year={2025}
}

@article{jin_agentreview_2024,
  title         = {{AgentReview}: Exploring Peer Review Dynamics with {LLM} Agents}, 
  author        = {Yiqiao Jin and Qinlin Zhao and Yiyang Wang and Hao Chen and Kaijie Zhu and Yijia Xiao and Jindong Wang},
  year          = {2024},
  archivePrefix = {arXiv},
  primaryClass  = {cs.CL},
  journal       = {arXiv preprint arXiv:2406.12708},
}

@article{sahu_reviewertoo_2025,
      title={{ReviewerToo}: Should {AI} Join The Program Committee?}, 
      author={Gaurav Sahu and Hugo Larochelle and Laurent Charlin and Christopher Pal},
      journal={arXiv preprint arXiv:2510.08867},
      year={2025}
}

@article{tian_scicode_2024,
      title={{SciCode}: A Research Coding Benchmark for Scientific Discovery}, 
      author={Yian Tian and Lijun Wu and Kevin Liu and Zecheng Zhang and Xun Liang and others},
      journal={arXiv preprint arXiv:2407.13168},
      year={2024}
}

@article{wang2025repomaster,
  title={Repomaster: Autonomous exploration and understanding of github repositories for complex task solving},
  author={Wang, Huacan and Ni, Ziyi and Zhang, Shuo and Lu, Shuo and Hu, Sen and He, Ziyang and Hu, Chen and Lin, Jiaye and Guo, Yifu and Chen, Ronghao and others},
  journal={arXiv preprint arXiv:2505.21577},
  year={2025}
}

@misc{chan_mlebench_2024,
  title={{MLE}-bench: Evaluating Machine Learning Agents on Machine Learning Engineering}, 
  author        = {Jun Shern Chan and Neil Chowdhury and Oliver Jaffe and James Aung and Dane Sherburn and Evan Mays and Giulio Starace and Kevin Liu and Leon Maksin and Tejal Patwardhan and Lilian Weng and Aleksander Mądry},
  year          = {2024},
  eprint        = {2410.07095},
  archivePrefix = {arXiv},
  primaryClass  = {cs.CL},
  journal={arXiv preprint arXiv:2410.07095},
}

@article{si2024can,
  title={Can llms generate novel research ideas? a large-scale human study with 100+ nlp researchers},
  author={Si, Chenglei and Yang, Diyi and Hashimoto, Tatsunori},
  journal={arXiv preprint arXiv:2409.04109},
  year={2024}
}

@article{lin_se-agent_2025,
      title={{SE-Agent}: Self-Evolution Trajectory Optimization in Multi-Step Reasoning with {LLM}-Based Agents}, 
      author={Jiaye Lin and Yifu Guo and Yuzhen Han and Sen Hu and Ziyi Ni and Licheng Wang and Mingguang Chen and Hongzhang Liu and Ronghao Chen and Yangfan He and Daxin Jiang and Binxing Jiao and Chen Hu and Huacan Wang},
      journal={arXiv preprint arXiv:2508.02085},
      year={2025}
}

@misc{arike2025tech,
      title={Technical Report: Evaluating Goal Drift in Language Model Agents}, 
      author={Rauno Arike and Elizabeth Donoway and Henning Bartsch and Marius Hobbhahn},
      year={2025},
      eprint={2505.02709},
      archivePrefix={arXiv},
      primaryClass={cs.AI},
      url={https://arxiv.org/abs/2505.02709}, 
}

@article{hu_controlled_2026,
      title={Controlled Self-Evolution for Algorithmic Code Optimization}, 
      author={Tu Hu and Ronghao Chen and Shuo Zhang and Jianghao Yin and Mou Xiao Feng and Jingping Liu and Shaolei Zhang and Wenqi Jiang and Yuqi Fang and Sen Hu and Huacan Wang and Yi Xu},
      journal={arXiv preprint arXiv:2601.07348},
      year={2026}
}

@article{luo2025llm4sr,
  title={Llm4sr: A survey on large language models for scientific research},
  author={Luo, Ziming and Yang, Zonglin and Xu, Zexin and Yang, Wei and Du, Xinya},
  journal={arXiv preprint arXiv:2501.04306},
  year={2025}
}

@article{xu2025can,
  title={Can LLMs Identify Critical Limitations within Scientific Research? A Systematic Evaluation on AI Research Papers},
  author={Xu, Zhijian and Zhao, Yilun and Patwardhan, Manasi and Vig, Lovekesh and Cohan, Arman},
  journal={arXiv preprint arXiv:2507.02694},
  year={2025}
}

@article{zhang2025evolving,
  title={The evolving role of large language models in scientific innovation: Evaluator, collaborator, and scientist},
  author={Zhang, Haoxuan and Li, Ruochi and Zhang, Yang and Xiao, Ting and Chen, Jiangping and Ding, Junhua and Chen, Haihua},
  journal={arXiv preprint arXiv:2507.11810},
  year={2025}
}

@article{zhang2026opennovelty,
  title={OpenNovelty: An LLM-powered Agentic System for Verifiable Scholarly Novelty Assessment},
  author={Zhang, Ming and Tan, Kexin and Huang, Yueyuan and Shen, Yujiong and Ma, Chunchun and Ju, Li and Zhang, Xinran and Wang, Yuhui and Jing, Wenqing and Deng, Jingyi and others},
  journal={arXiv preprint arXiv:2601.01576},
  year={2026}
}

@article{zhang_noveltybench_2025,
      title={{NoveltyBench}: Evaluating Creativity and Diversity in Language Models}, 
      author={Yiming Zhang and Harshita Diddee and Susan Holm and Hanchen Liu and Xinyue Liu and Vinay Samuel and Barry Wang and Daphne Ippolito},
      journal={arXiv preprint arXiv:2504.05228},
      year={2025}
}

@article{baek_researchagent_2025,
      title={{ResearchAgent}: Iterative Research Idea Generation over Scientific Literature with Large Language Models},
      author={Jinheon Baek and Sujay Kumar Jauhar and Silviu Cucerzan and Sung Ju Hwang},
      journal={Proceedings of the 2025 Conference of the Nations of the Americas Chapter of the Association for Computational Linguistics: Human Language Technologies (Volume 1: Long Papers)},
      pages={6709--6738},
      year={2025}
}

@article{wu_autogen_2023,
  title={Autogen: Enabling next-gen llm applications via multi-agent conversation framework},
  author={Wu, Qingyun and Bansal, Gagan and Zhang, Jieyu and Wu, Yiran and Zhang, Shaokun and Zhu, Erkang and Li, Beibin and Jiang, Li and Zhang, Xiaoyun and Wang, Chi},
  journal={arXiv preprint arXiv:2308.08155},
  year={2023}
}

@article{hong_metagpt_2023,
      title={{MetaGPT}: Meta Programming for Multi-Agent Collaborative Framework}, 
      author={Sirui Hong and Xiawu Zheng and Jonathan Chen and Yuheng Cheng and Jinlin Wang and Ceyao Zhang and Zili Wang and Steven Ka Shing Yau and Zijuan Lin and Liyang Zhou and others},
      journal={arXiv preprint arXiv:2308.00352},
      year={2023}
}

@article{chai2025scimaster,
  title={SciMaster: Towards General-Purpose Scientific AI Agents, Part I. X-Master as Foundation: Can We Lead on Humanity's Last Exam?},
  author={Chai, Jingyi and Tang, Shuo and Ye, Rui and Du, Yuwen and Zhu, Xinyu and Zhou, Mengcheng and Wang, Yanfeng and Zhang, Yuzhi and Zhang, Linfeng and Chen, Siheng and others},
  journal={arXiv preprint arXiv:2507.05241},
  year={2025}
}

@article{zeng2025glm,
  title={Glm-4.5: Agentic, reasoning, and coding (arc) foundation models},
  author={Zeng, Aohan and Lv, Xin and Zheng, Qinkai and Hou, Zhenyu and Chen, Bin and Xie, Chengxing and Wang, Cunxiang and Yin, Da and Zeng, Hao and Zhang, Jiajie and others},
  journal={arXiv preprint arXiv:2508.06471},
  year={2025}
}

@article{team2025gemma,
  title={Gemma 3 technical report},
  author={Team, Gemma and Kamath, Aishwarya and Ferret, Johan and Pathak, Shreya and Vieillard, Nino and Merhej, Ramona and Perrin, Sarah and Matejovicova, Tatiana and Ram{\'e}, Alexandre and Rivi{\`e}re, Morgane and others},
  journal={arXiv preprint arXiv:2503.19786},
  year={2025}
}

\clearpage
\newpage
\beginappendix

\end{document}